\documentclass[conference]{IEEEtran}
\IEEEoverridecommandlockouts
\IEEEpubid{\makebox[\columnwidth]{ 979-8-3503-5067-8/24/\$31.00~\copyright2024 IEEE \hfill}
\hspace{\columnsep}\makebox[\columnwidth]{ }}

\usepackage{romannum}
\usepackage{cite}
\usepackage[utf8]{inputenc}
\usepackage[T1]{fontenc}
\usepackage{amsmath,amssymb,amsfonts}
\usepackage{algorithmic}
\usepackage{graphicx}
\usepackage{textcomp}
\usepackage{xcolor}
\usepackage{float}
\def\BibTeX{{\rm B\kern-.05em{\sc i\kern-.025em b}\kern-.08em
    T\kern-.1667em\lower.7ex\hbox{E}\kern-.125emX}}
\begin{document}

\title{Project Beyond: An Escape Room Game in Virtual Reality to Teach Building Energy Simulations\\

\thanks{This work was supported by a grant from the Austrian Research Promotion Agency (FFG) program \textit{Stadt der Zukunft}, project number FO999887002.}
}

\author{
\IEEEauthorblockN{
Georg Arbesser-Rastburg\IEEEauthorrefmark{1}, 
Saeed Safikhani\IEEEauthorrefmark{2}, 
Matej Gustin\IEEEauthorrefmark{3}, 
Christina Hopfe\IEEEauthorrefmark{4}, 
Gerald Schweiger\IEEEauthorrefmark{5}
}
\IEEEauthorblockA{
\textit{Graz University of Technology}, Graz, Austria \\
Email: 
\IEEEauthorrefmark{1}georg.arbesser-rastburg@tugraz.at,
\IEEEauthorrefmark{2}s.safikhani@tugraz.at,
\IEEEauthorrefmark{3}m.gustin@tugraz.at,\\
\IEEEauthorrefmark{4}c.j.hopfe@tugraz.at,
\IEEEauthorrefmark{5}gerald.schweiger@tugraz.at
}

\IEEEauthorblockN{
\\Johanna Pirker
}
\IEEEauthorblockA{
\textit{Ludwig-Maximilians-Universität München} \
Munich, Germany \\
Email: jpirker@iicm.edu
}
}

\maketitle
\IEEEpubidadjcol

\begin{abstract}
In recent years, Virtual Reality (VR) has found its way into different fields besides pure entertainment. One of the topics that can benefit from the immersive experience of VR is education. Furthermore, using game-based approaches in education can increase user motivation and engagement. Accordingly, in this paper, we designed and developed an immersive escape room game in VR to teach building energy simulation topics. In the game, players must solve puzzles like, for instance, assembling walls using different materials. We use a player guidance system that combines educational content, puzzles, and different types of hints to educate the players about parameters that influence energy efficiency, structural resistance, and costs. To improve user onboarding, we implemented a tutorial level to teach players general interactions and locomotion. 

To assess the user experience, we evaluate both the tutorial and the game with an expert study with gaming and VR experts (n=11). The participants were asked to play both the tutorial level and the escape room level and complete two sets of post-questionnaires, one after the tutorial and one after the puzzle level. The one after the tutorial level consisted of NASA-TLX and SUS questionnaires, while after the escape room level we asked users to complete the NASA-TLX, UESSF, and PXI questionnaires. The results indicate that the onboarding level successfully provided good usability while maintaining a low task load. On the other hand, the escape room level can provide an engaging, visually appealing, and usable learning environment by arousing players' curiosity through the gameplay. This environment can be extended in future development stages with different educational contents from various fields.   
\end{abstract}

\begin{IEEEkeywords}
Virtual Reality, Escape Room, Gamification, Educational Game
\end{IEEEkeywords}

\section{Introduction} \label{introduction}
\begin{figure*}
    \centering
    \includegraphics[width = \linewidth]{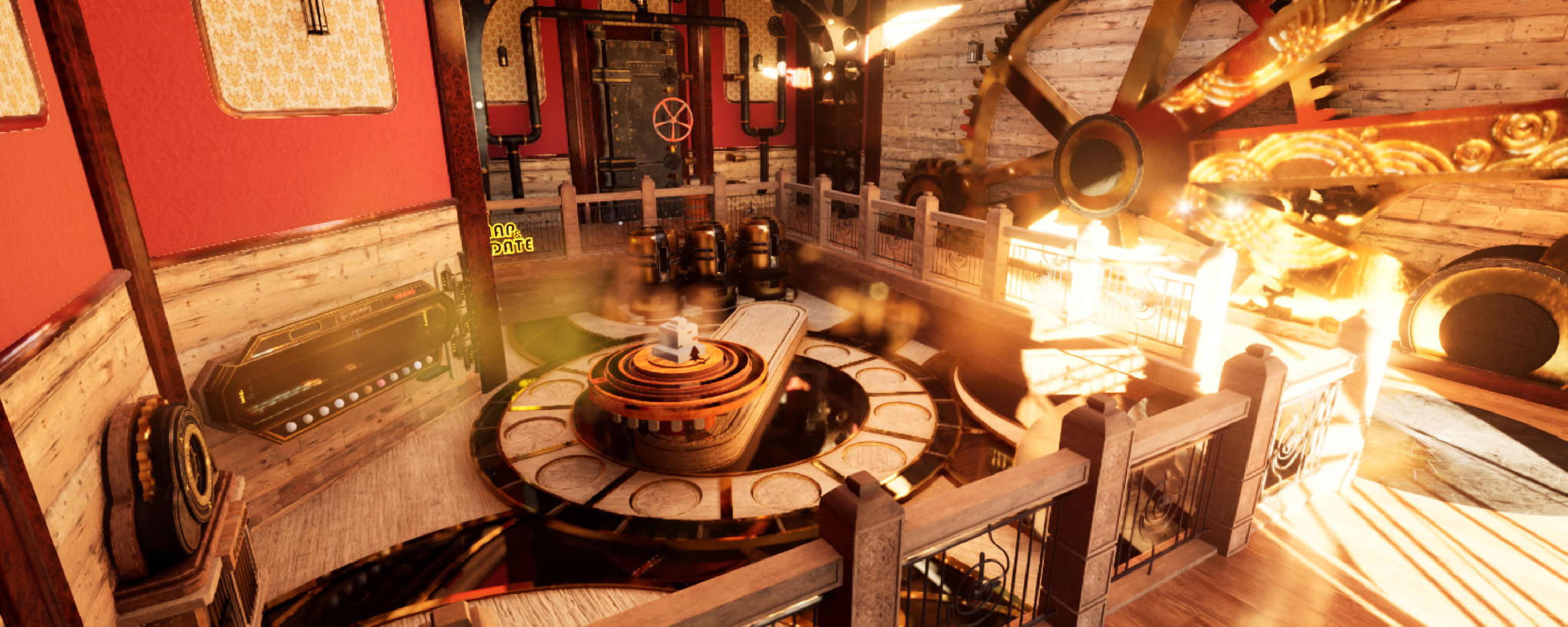}
    \caption{An overview of the escape room in Project Beyond.}
    \label{fig:beyond_overview}
\end{figure*}
In recent years, an ever-growing interest in interactive, student-centered approaches that shift from the classical paradigm of teacher-centered education could be observed. Students now have different expectations regarding educational methods, preferring to learn actively with an approach that is based on making experiences rather than listening passively, after having grown up in an ever more digital world \cite{Oblinger2004TheEngagement}.

Combining games with educational content is one way of meeting this demand for interactivity. This is done to increase the learner's engagement and motivation by evoking emotions like enjoyment or accomplishment during gameplay \cite{Bray2023EvaluationProtocol}. Such games are often called \textit{serious games}. A serious game is a game that not only aims to entertain the players but also tries to fulfill an additional goal, such as educating them\cite{Dorner2016SeriousPractice}. 

Numerous approaches to serious gaming have been taken in the past, for instance, for medical training or architecture education. Studies have shown that serious games for educational purposes positively influence user engagement and cognitive abilities while helping the players keep a positive attitude \cite{Zhonggen2019ADecade}.

Escape room games are one possible approach to educational games. In an escape room game, players must solve puzzles or complete tasks to reach a specific goal, such as escaping the room where the game takes place within a specified time limit \cite{Nicholson2015PeekingFacilities}. Typically, this is a team-based activity where players collaborate to find clues and solve puzzles.

While most escape rooms serve recreational purposes, they have also been applied as a teaching tool in different fields, for instance, mathematics \cite{Fuentes-Cabrera2020LearningStudy, Rosillo2021EscapeEra}, healthcare \cite{Guckian2020TheEducation, Friedrich2019EscapingCurriculum}, computer science \cite{Beguin2019Computer-Security-OrientedRoom, Ho2018UnlockingClassroom} or chemistry \cite{Avargil2021EducationalEscape, Vergne2019EscapeExperiment}, allowing students to explore a specific topic interactively. In an educational context, escape rooms are used to foster team collaboration, critical thinking, and creativity and to create quasi-realistic scenarios while providing fun to the players \cite{Taraldsen2022AVoid}.

However, classical educational escape rooms have several limitations in practice, as they require a (class-)room that has to be specifically prepared for the escape room experience, which also leads to the need for a sufficient budget and personnel \cite{Fotaris2019EscapeReview}. Hence, a number of digital escape room games have been developed to mitigate these issues. As previous studies have shown, digital escape rooms can be a viable option to help students understand scientific content \cite{Ang2020PhysicalBonding, Monnot2020NewParticipation} while increasing their feeling of autonomy and creativity \cite{Pozo-Sanchez2022ComparingRooms}. 

A particular type of digital escape room uses virtual reality (VR) to immerse players in the experience using Head Mounted Displays (HMD). In recent years, VR devices have been used for various applications, from classic entertainment and various industrial applications \cite{Damiani2018AugmentedEra} to different educational approaches \cite{Radianti2020AAgenda}. In general, using VR for educational purposes has several advantages, for instance, when training scenarios that are hard to replicate in the real world or that are potentially dangerous. In addition, using VR with HMDs can lead to increased motivation \cite{Freina2015APerspectives}. One niche in VR education is the aforementioned use of digital escape room games. A few examples of such games can be found in the literature. 

In \cite{Yeasmin2020ImplementationGame} an escape room game in VR is described that consists of mathematical, pattern recognition, and combination puzzles. The study aimed to evaluate the influence of teleportation on cybersickness.
Another example is a two-player VR game where players have to solve puzzles while being limited as to how they can communicate with each other during the experience \cite{Ioannou2023CollaborativeRooms}. Not being able to communicate via voice or text, players resorted to innovative methods of communication, which led to increased user engagement.
Kiruthika et al. \cite{Kiruthika2022VirtualRoom} describe a VR escape room game consisting of two levels (a seaport and a palace), where users have to solve different memory and logic puzzles.
A VR escape room game described in \cite{David2019DevelopmentTechnology} was developed for Samsung Gear VR and contains three different logic puzzles that must be solved sequentially.

While the aforementioned games can be categorized as recreational escape room games, some examples of educational escape room games can also be found.
A VR escape room game that can be played by individuals or by teams is presented in \cite{Staneva2023GamificationTechnologies}. Players must escape from an ancient tomb through several levels, including learning material such as text, video, or audio.
In the game proposed in \cite{Elford2021StereoisomersReality}, students have to solve puzzles related to stereochemistry while trying to join a fictional secret intelligence agency. The goal was to foster the student's knowledge of stereochemistry and the development of soft skills.
In \cite{Christopoulos2023EscapingEducation}, a VR escape room game is presented that deals with different aspects of enzymes for biology education. Their results indicate that the game increased short-term memory of the conveyed content compared to an educational video while not improving long-term knowledge.

Another area that makes use of innovative interactive applications in an educational context is civil engineering. For instance, \cite{Ebner2007SuccessfulEngineering} describe a game that teaches students about calculating internal forces static determinate systems. Whisker et al. showcase two VR environments, one used to display 4D CAD models for educational purposes, the other being a virtual construction project \cite{Whisker2003UsingEducation}. However, we could not find any escape room games for training civil engineering students.

To close the gap between educational VR escape room games and civil engineering, we present such a game in this paper (Section \ref{beyond}). 
This game targets civil engineering students and deals with the intricacies of building energy simulations and wall constructions. Students must solve various puzzles to escape a locked room. We use a carefully crafted player guidance system that combines educational content, puzzles, and different types of hints to 
i) provide educational context and information for the puzzles the players have to solve, ii) describe important issues and pitfalls frequently encountered in civil engineering education and iii) guide players through the escape room experience step-by-step.

To evaluate our system's usability, the resulting task load, and user experience, we ran an expert study with a group of game developers and VR experts (Section \ref{evaluation}). Finally, after presenting our findings and discussing current limitations, we outline directions for future research (Section \ref{discussion}).

\section{The Project Beyond} \label{beyond}
To implement our educational content on building energy simulation in a gamified environment, we developed an escape room game in VR. In the beginning, players find themselves locked in a room and have to solve puzzles related to building energy simulations and wall constructions. To leave the room, they have to solve several quests that combine puzzles with educational content to unlock the door's four locks.

\subsection{Tackling Common Issues in Civil Engineering Education}
Applying educational content learned through traditional lectures to real-world scenarios can be challenging because of potential mistakes or misunderstandings. 
The main story of our escape room game is inspired by common issues that students report during their energy simulation education. The following issues were frequently encountered during classes: forgetting to consider a building's location or orientation, incorrect wall layer assembly, and mistakes related to wall layer properties.
Based on these challenges, we designed corresponding puzzles in the game. We expect that tackling these issues in a gamified environment helps students understand the concepts deeper and reduces their future design mistakes.

\subsection{Educational Concept}
We rely on a close connection between visual and audio hints, quests, and gadgets visualizing simulation results to educate players about the various building and building energy simulation concepts in the game. The game is structured into several quests, each dealing with one aspect, like considering the correct time of day/month and location while planning a building or correctly assembling different wall constructions.

We follow the same procedure for all topics included in the game: First, we describe the general problem players must deal with and provide educational context via visual hints and audio voice-over (e.g., the importance of a building's orientation). Next, players must find cues (e.g., a map detailing the correct orientation). With this information, players can solve the quest and are immediately notified via audio and visual feedback.

While setting the time of day/month and location mostly requires players to find clues in the room and correctly use this information to proceed to the next quest, assembling wall consturctions is a more complex procedure. Here, players must balance several counter-acting simulation results and iteratively optimize their wall construction. For instance, a wall with high structural stability will cost more, while a wall with thinner insulation will lead to lower costs but might show mold on the wall. This enables players to learn about these connections and teaches them to carefully balance the individual layers while getting constant feedback about their decisions and performance.

\subsection{Technology}
We developed our VR experience in Unreal Engine 5 (UE)\footnote{www.unrealengine.com}. UE helped us to design a realistic environment to improve the sense of presence in VR \cite{Schwind2019UsingReality}. 
For the implementation of VR interactions, we used the OpenXR plugin\footnote{https://docs.unrealengine.com/5.3/en-US/developing-for-head-mounted-experiences-with-openxr-in-unreal-engine/} in UE to support a wide range of VR devices (e.g., SteamVR, Oculus, and Windows Mixed Reality).

\subsection{Interaction and Locomotion}
Based on the OpenXR API, we developed a custom VR character to support the required functionality for our gameplay. This character supports two different modes of locomotion in addition to physical movement: 1) Teleportation and 2) Smooth Movement. User can freely choose between them according to their preferences and gameplay. The teleportation system can reduce the possibility of motion sickness, but might lead to disorientation. On the other hand, smooth movement has less chance of disorientation and is easier to control, but might cause higher motion sickness. 
In order to provide a comprehensive set of gameplay elements, we implemented different interaction systems. The main interactions in Project Beyond are "grab" and "placement". Grabbing items can be done with one or two hands based on the object type leading to different behaviors, such as item pick-up or object rotation. In addition to the grab interaction, we introduced pressure-dependent and hover-interactable objects. The pressure-dependent objects use UE's physics constraints to simulate a realistic behavior, but the hover-interactable object triggers the interactions just by overlapping a trigger area.

\subsection{Tutorial Level}
\begin{figure}
    \centering
    \includegraphics[width = 8.5cm]{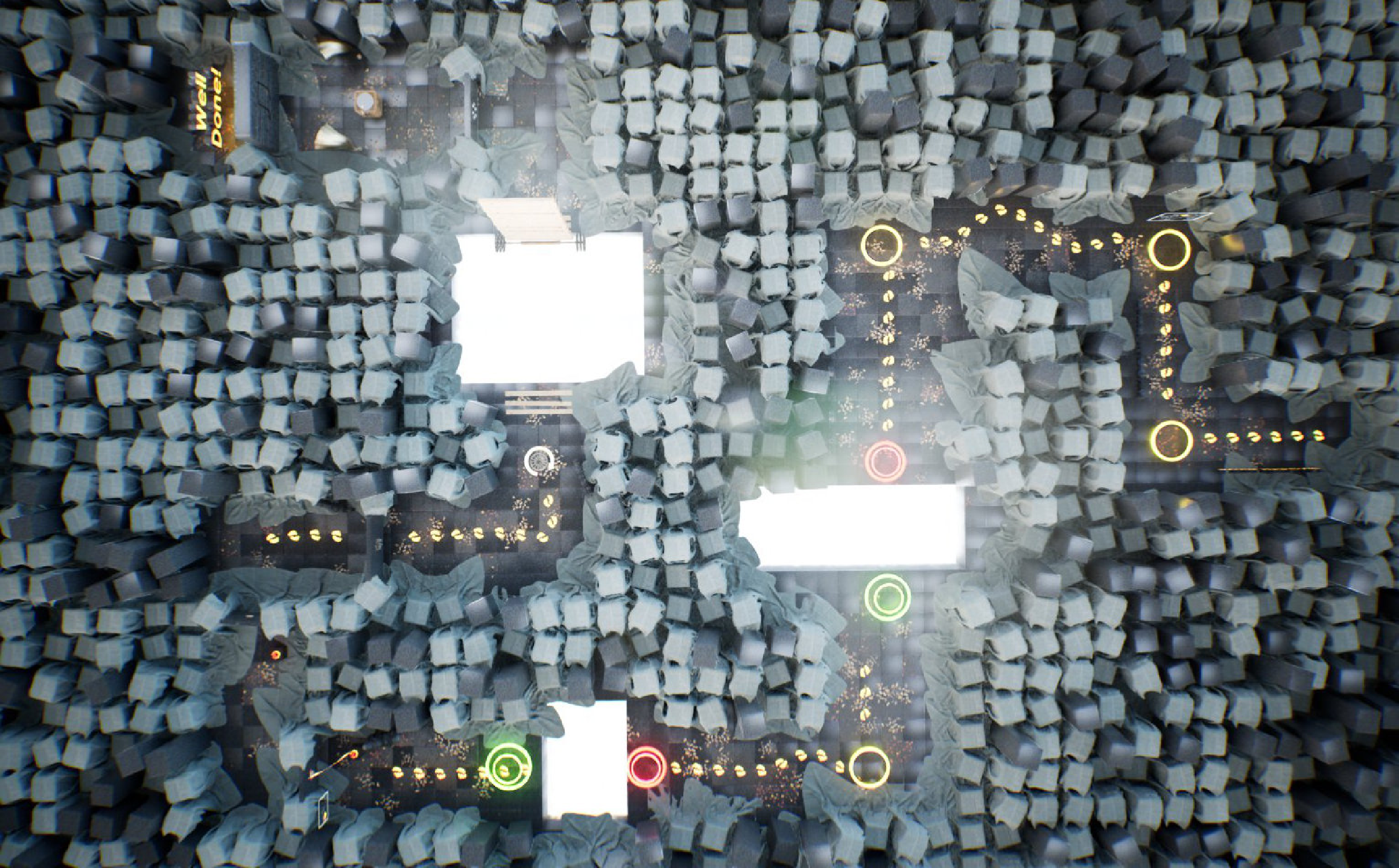}
    \caption{Top view of the onboarding level with hints to learn different interactions, e.g., golden footprints indicate smooth movement paths.}
    \label{fig:onboarding_topview}
\end{figure}

We implemented a dedicated tutorial level for players to get acquainted with VR in general, locomotion, and possible interactions (see Fig.~\ref{fig:onboarding_topview}). This should prevent new users from getting overwhelmed in the escape room by having to find clues and solve puzzles while learning to move and interact in VR simultaneously. We decided to separate the tutorial from the actual escape room for two reasons:
\begin{enumerate}
    \item Some players might already be familiar with VR and might want to skip the tutorial. Since the tutorial is separate, players can choose whether they want to complete it or not.
    \item Players should only learn about locomotion and VR interactions in an abstract manner. The tutorial should not give away any hints or influence the players with regard to the escape room game in any way.
\end{enumerate} 
In the tutorial, we cover interactions for the locomotion system, grab, physical touch, and rotatable objects. For each interaction, we use different symbols, indicators, and controller schema. For example, we use footprint symbols to indicate the path for smooth movement and a controller figure highlighting the required button to move toward the path's target location.
To simplify the procedure of level design for the onboarding level, we used Procedural Content Generation (PCG)\footnote{https://docs.unrealengine.com/5.3/en-US/procedural-content-generation--framework-in-unreal-engine/} in UE.
PCG helps to iterate through design ideas with ease and facilitates future level extensions if needed.

\subsection{Quest and Hint System}
A guidance system was required to include educational content and describe special interactions or machines in the room. Furthermore, players need to be informed about their progress and receive feedback on whether they are on the correct path. Considering these points, we developed a special quest and hint system to guide players throughout the whole game. 
Each quest in the game may contain several hints according to the task. The hints can be found in the level as pieces of paper with text and a figure on them. As reading text in VR can be difficult depending on the HMD's quality, and therefore, reading a long text is not practical in VR, we included a voice-over for each hint to describe the hint in more detail.
In addition, players may need to review hints again after collecting them. To enable this, we implemented two devices in the game to review previously found hints. A projector device projects the content of a paper hint onto a curtain in a bigger size, which allows players to repeat the hint and makes it easier to read. In addition, a cassette player is attached to the player's left hand and can repeat the voiceover corresponding to the paper hint selected at the projector.

Hints are spawned when a quest gets activated. This allows us to closely link the hints to the player's progress and make sure that only relevant hints are visible to the player. To indicate to the players when they complete a quest, an audio/visual feedback system is used. When completing a smaller quest, a short chime is played. When a more important quest is solved, i) a longer sound is played, ii) the light in the room runs through a complete day/night cycle with natural sunlight during the day and artificial light at night, and iii) one lock of the door is unlocked.

\subsection{Room Structure and Machines}
\begin{figure}
    \centering
    \includegraphics[height = 5cm]{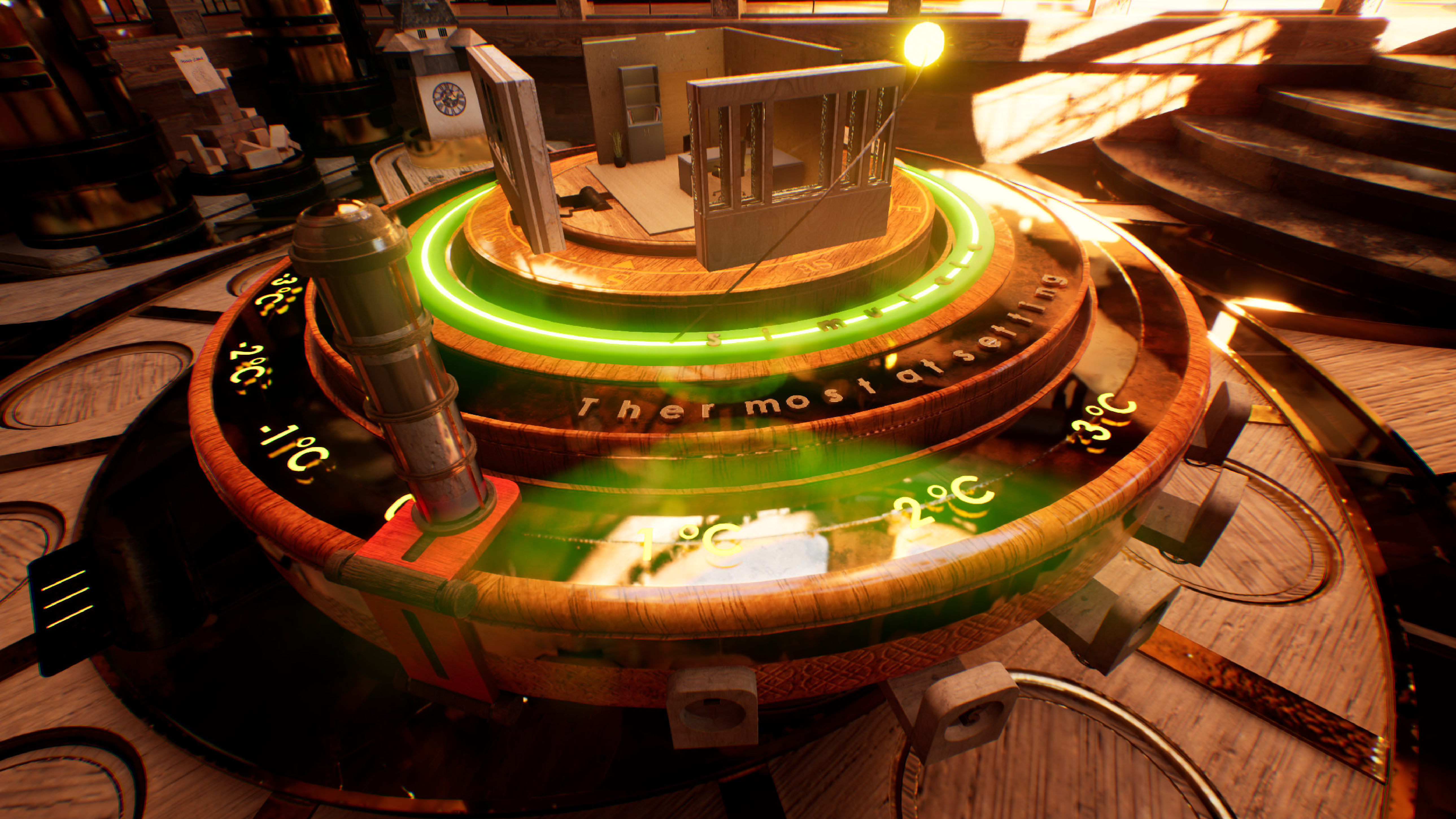}
    \caption{The building simulation desk showing the applied wall materials on the exterior walls of the simulation room.}
    \label{fig: simulation_desk}
\end{figure}

The escape room in Project Beyond can be divided into four main areas: 1) hint review (projector), 2) building simulation desk, 3) wall assembly, and 4) result visualizations.
The projector area contains the projector and a curtain to visualize the hints.
The building simulation desk is the center area in the game around which the main story is built (see Fig. ˜\ref{fig: simulation_desk}). It consists of three layers, of which the two lower ones can be rotated. Turning these layers can change different parameters, such as time, month, or various simulation settings, during the gameplay. Initially, a building is shown on the top layer. Later in the game, this building is replaced with a representation of an office room. In addition, there is a model of the sun to indicate the time of day and the influence of sunlight on the room based on location, month, and time. At a later stage in the game, when players create a wall section, they can use the desk to assign the wall section to the office room. This assignment leads to an energy simulation for the room, and users can see the result as gadgets.

The wall section assembly consists of three devices: i) information center, ii) section spawner, and iii) section assembly (left wall in Fig.~\ref{fig:beyond_overview}).
The information center is responsible for displaying the order of wall layers for a structural material type (masonry, reinforced concrete, or timber). Players can spawn layers and set their thickness in the section spawner. Finally, they can use the wall assembly device to assemble layers based on the required layer order. The result of these three devices can then be converted into a wall sample that can be assigned to the walls of the sample room. 

The result visualization area is the ring around the building simulation desk and contains placeholders for several gadgets. These gadgets visualize results regarding the building energy simulation of the sample room and other performance indicators related to the wall construction.

\subsection{Simulation API}
The room model on the table is based on an office at Graz University of Technology. Based on the players' inputs, the room's annual energy consumption for heating and cooling is calculated. Since this calculation with regular building energy simulation tools takes too long for an interactive game, a machine learning model is employed instead. The model was trained with about 12,000 building energy simulation runs to enable a swift calculation for the VR environment. 
The calculation was outsourced to a server that provides an interface for starting calculations and retrieving simulation results. After retrieving the results, several gadgets show an evaluation of the wall construction and the simulation parameters.

\begin{figure}
    \centering
    \includegraphics[width = 8.5cm]{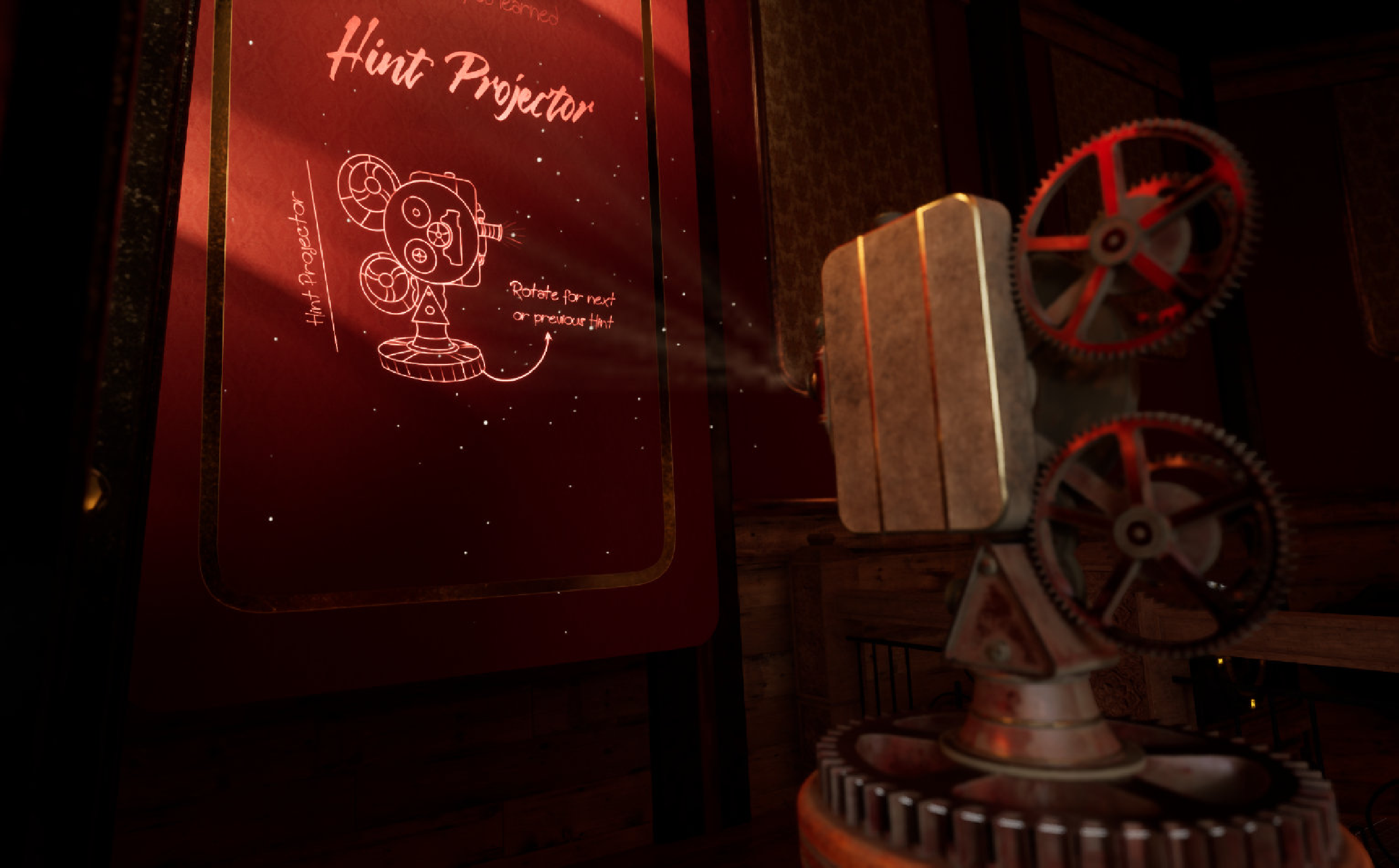}
    \caption{The projector can be used to review the collected paper hints on a bigger scale.}
    \label{fig: projector}
\end{figure}

\subsection{Gadgets}
\begin{figure}
    \centering
    \includegraphics[height = 5cm]{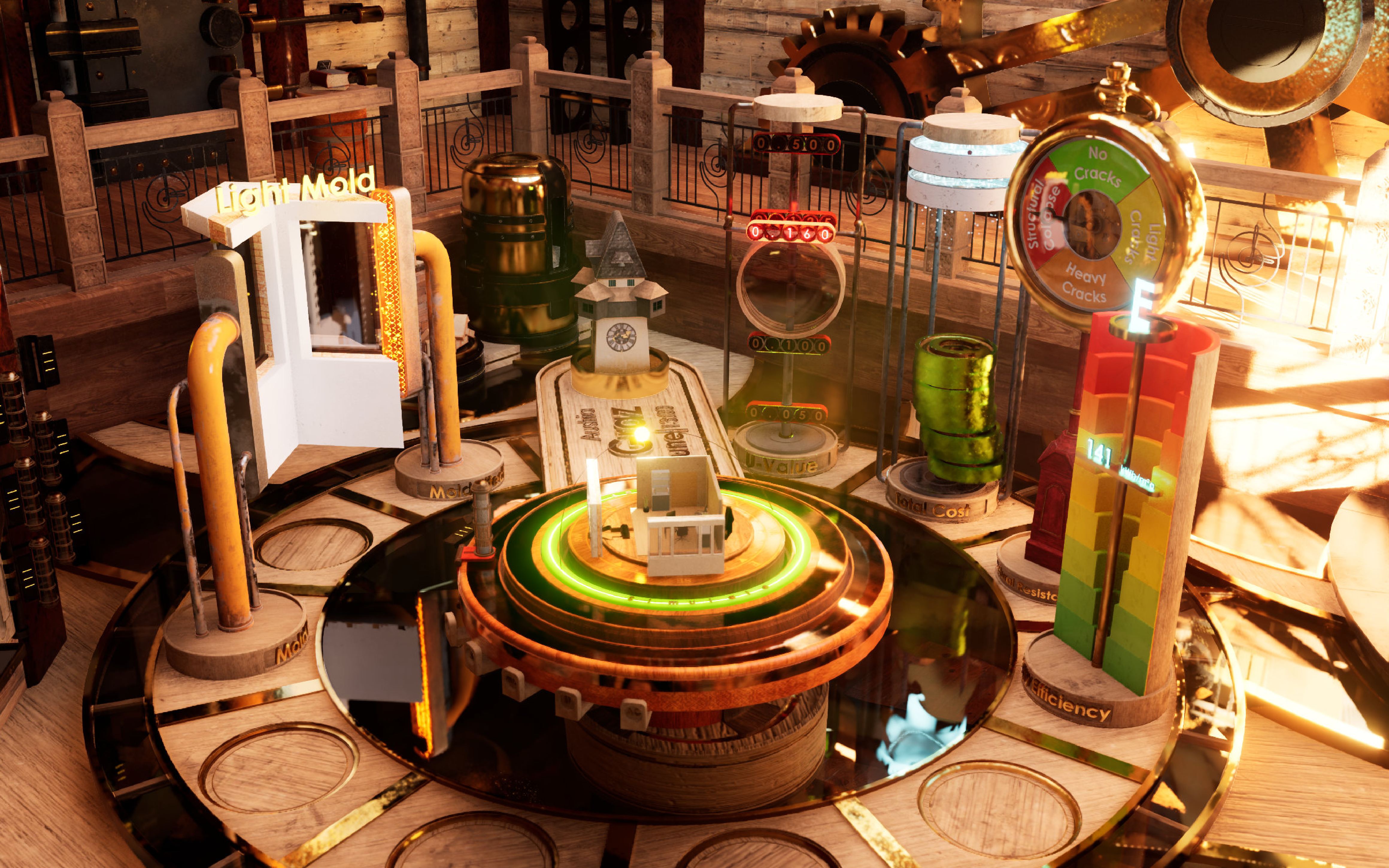}
    \caption{The simulation result gadgets (from left to right): mold, U-value, costs, structural stability, and energy efficiency.}
    \label{fig: gadgets}
\end{figure}

To evaluate players' choices regarding wall assembly, we implemented five gadgets representing i) mold on the walls, ii) the walls' U-value (rate of heat transfer through the walls), iii) wall construction costs, iv) the wall's structural stability, and v) the room's energy efficiency (see Fig. ˜\ref{fig: gadgets}). The mold gadget shows whether the wall is susceptible to mold using a rough estimation based on the wall's dew point calculation, from no mold to heavy mold, by showing the mold visually on the wall and displaying text above the wall section. The U-value indicator is a slider that shows the U-value in red or green, depending on whether the value is in a pre-defined acceptable range. The costs are calculated based on each layer's thickness and material and are represented with green, gold, and red coins (corresponding to cheap, medium-cost, and expensive walls). The structural stability gadget demonstrates the level of possible cracks in the wall section, from no cracks to complete structural collapse. Structural stability is again estimated based on layer thicknesses and materials. Finally, the energy efficiency gadget visualizes the room's overall energy efficiency based on the wall construction and several additional parameters regarding the room's operation and layout (i.e., using cooling, window shades, or changing the set point temperature as well as the window's U-value and solar heat gain coefficient). Energy efficiency is shown with a rating from A+ to H on top and the room's estimated energy consumption as a slider.

\section{Evaluation} \label{evaluation}
As a first step towards validating our game, we conducted a study with VR and game development experts to evaluate the general usability of the onboarding level and the game itself. 

\subsection{Methodology}
The study comprised one pre-questionnaire and two post-questionnaires, one completed after the onboarding level (post-questionnaire I) and one after playing the escape room game (post-questionnaire II). We split the evaluation into two parts since we expected very different results between the two levels, especially concerning task load. The questionnaires were provided via a LimeSurvey instance. Eleven VR and gaming experts -- bachelor, master, and PhD computer science students -- participated in the study.

\textbf{Material:} The pre-questionnaire contained only one question: "How much prior experience do you have in VR environments?" where the participants were asked to provide a short free text answer.
Post-questionnaire I consisted of the \textit{NASA Task Load Index} (NASA-TLX) questionnaire \cite{Hart1988DevelopmentResearch} to measure the task load while completing the tutorial and the \textit{System Usability Scale} (SUS) \cite{Brooke1996SUS:Scale} to evaluate its usability. Post-questionnaire II comprised the NASA-TLX, \textit{User Engagement Scale Short Form} (UESSF) \cite{OBrien2018AForm} to measure user engagement, and the \textit{Player Experience Inventory} (PXI) \cite{Abeele2020DevelopmentConsequences} to evaluate the player experience. 

\textbf{Setup:} For running the virtual environment, we used a PC (Intel Core i5-13600K, 32 GB RAM, RTX 3080 Ti) together with a Meta Quest 2\footnote{https://www.meta.com/at/en/quest/products/quest-2/} connected wirelessly via Air Link\footnote{https://www.meta.com/en-gb/help/quest/articles/headsets-and-accessories/oculus-link/connect-with-air-link/}. The study took place in a room with about 4m x 6m as a playable area. During the study, one or two supervisors were in the room to help participants connect the HMD, start the game, and guide them through the study. Supervisors could follow the participants' actions on a TV connected to the PC.

\textbf{Procedure:} We ran the study one participant at a time. We first verbally introduced Project Beyond to the participants and provided a short written description of key terminology related to building energy simulations and wall structure. This was deemed necessary as the game is targeted at civil engineering students who have basic domain knowledge. In contrast, the computer scientists were mostly unfamiliar with specific terms, especially not in English. 

\begin{figure}[htb]
    \centering
    \includegraphics[width = 8.5cm]{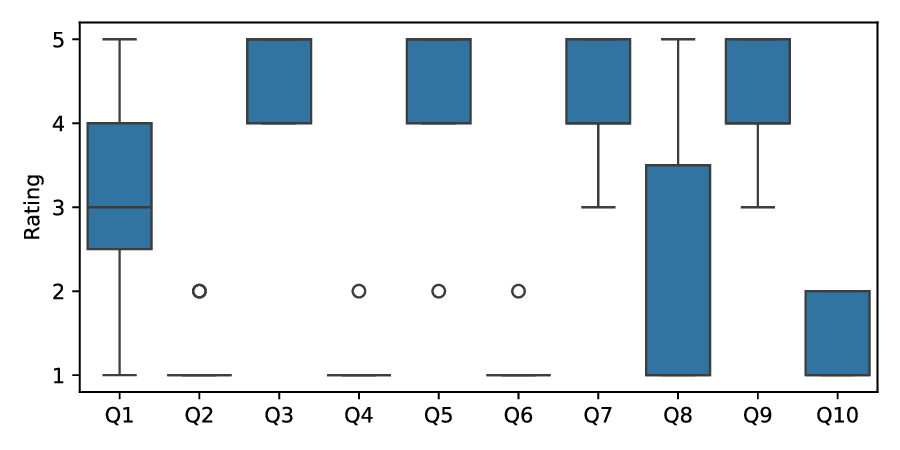}
    \caption{SUS results of the tutorial level.}
    \label{fig: sus}
\end{figure}

After this introduction, the participants completed the pre-questionnaire and played the complete tutorial level. During the tutorial, participants had to follow four paths for smooth movement and two sections for teleportation locomotion. Afterward, the participants could choose between either locomotion system and had to complete two grabbing tasks, open two doors by pressing buttons, rotate a wheel to lower a draw bridge, and lower a pickup onto a vinyl record. This was followed by the participants completing post-questionnaire I.

Next, the participants played the actual escape room. They were allowed to ask for three hints from the supervisors during the game and were given about 45 minutes to play. Players followed the storyline from learning to use the cassette player and hint projector through setting the correct building orientation, time of day and month, and location to assembling one wall. Participants were not asked to optimize the wall since this requires more profound domain knowledge and was deemed too time-expensive for this study. Finally, the participants were asked to complete post-questionnaire II.

\subsection{Results}
The results of the SUS questionnaire (Fig.~\ref{fig: sus}) show very good usability for the tutorial level (AVG=84.3, SD=9.09). The average value for Q1 ("I think that I would use this system frequently") is below the acceptable range. Verbal feedback from the participants indicates that they did not believe they would need to play the tutorial again, as the interactions were clear and easy for them, which could explain this value. The result of Q8 ("I found the system very cumbersome to use") shows a great spread in responses that could be due to some participants misunderstanding the question.

\begin{figure}[t]
    \centering
    \includegraphics[width = 8.5cm]{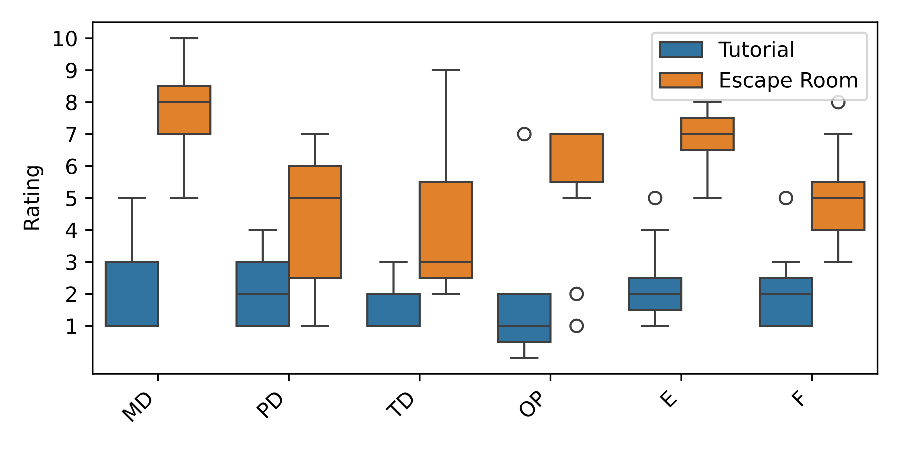}
    \caption{Comparison of NASA-TLX results of the tutorial and the escape room levels. MD: Mental Demand, PD: Physical Demand, TD: Temporal Demand, OP: Overall Performance: E: Effort, F: Frustration.}
    \label{fig: tlx}
\end{figure}

The task load evaluation using NASA TLX questionnaires shows a very low demanding experience in the onboarding level (Fig.~\ref{fig: tlx}). Furthermore, users found themselves successful in fulfilling the tutorial tasks (AVG=1.55, SD=1.97). The overall experience of the onboarding level was not frustrating (AVG=2.09, SD=1.22), and they did not need extensive effort to finish this level (AVG=2.27, SD=1.27). The outcome of this questionnaire for the main puzzle level is totally different from the onboarding level. As we expected, based on the features of an escape room game, players experienced a higher mental demand (tutorial: AVG=2.45, SD=1.37; escape room: AVG=7.73, SD=1.42) and effort (AVG=6.91, SD=0.94) to solve the puzzles. The average frustration level in the escape room level is medium (AVG=5.00, SD=1.15). The temporal demand was below the average (AVG=4.09, SD=2.21) as we did not consider any time pressure for the gameplay. Although the interactions are similar to the tutorial level, participants reported higher physical demand in the puzzle level (tutorial: AVG=2.09, SD=1.04; escape room: AVG=4.45, SD=2.11), which could be due to the longer duration of the experience. Finally, players did not find themselves successful in solving puzzles of the escape-room level (AVG=5.64, SD=2.16).  

The user engagement questionnaire shows a high value of focused attention (FA: AVG=71.2, SD=24.6). This value can be interpreted as feeling high absorption in the interaction and losing track of time. The observed average value of perceived usability is lower than the medium (PU: AVG=41.6, SD=26.7). This can be interpreted as an acceptable degree of control and effort. The attractiveness and visual appeal of the game are perceived very well and accepted with very high values (AE: AVG=92.4, SD=10.9). The results show users interested in the interactive tasks and willing to recommend the application to others (RW: AVG=83.3, SD=15.3).

\begin{figure}[htb]
    \centering
    \includegraphics[width = 8.5cm]{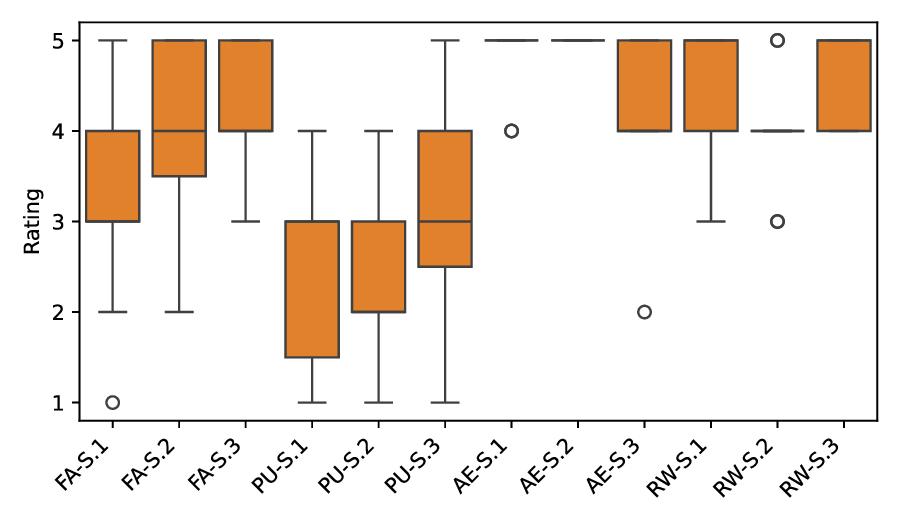}
    \caption{UESSF results of the escape room level.}
    \label{fig: uessf}
\end{figure}

Based on the results of the PXI questionnaires, players found the game meaningful overall (Meaning: AVG=1.15, SD=1.23). Furthermore, the game was highly successful in arousing the users' curiosity (Curiosity: AVG=2.06, SD=1.14). The mastery and autonomy items are slightly lower than the medium value (Mastery: AVG=-0.2, SD=1.7, Autonomy: AVG=-0.06, SD=1.8). The immersion and audio/visual presentation were perceived very well in the game with high scores (Immersion: AVG=1.88, SD=1.3, Audio/Visual: AVG=2.55, SD=0.93). The progress feedback was slightly above the medium score (Progress Feedback: AVG=0.64, SD=1.57). Users reported that the challenges in the game were a bit difficult to handle (Challenges: AVG=-0.12, SD=1.58). However, results show that the game was mainly easy to control (Ease of Control: AVG=1.18, SD=1.65). The goal of the game was grasped in general with an acceptable score (Goals and Rules: AVG=0.97, SD=1.4), and players highly enjoyed the game (Game Enjoyment: AVG=2.09, SD=0.99).

\begin{figure}[t]
    \centering
    \includegraphics[width = 8.5cm]{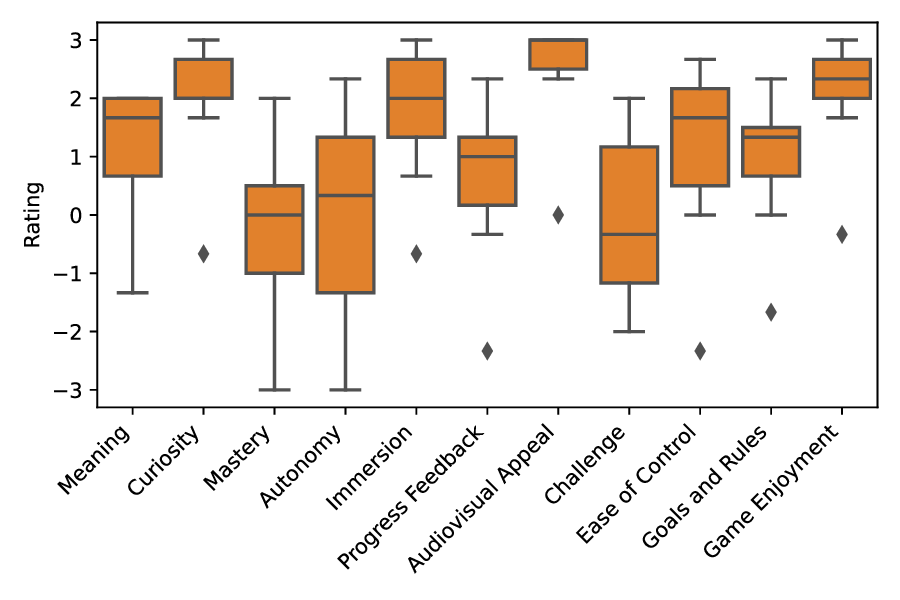}
    \caption{PXI results of the escape room level.}
    \label{fig: pxi}
\end{figure}
\section{Discussion} \label{discussion}
In this study, we aim for two main goals when evaluating the functionality of our VR environment and its player guidance and education system. First, we want to determine whether our onboarding level can provide acceptable usability while preserving a low task load. This is important to avoid overwhelming users before entering the main game. Second, we are investigating the user experience in our puzzle environment to figure out whether the game can increase user engagement and curiosity to discover the educational content of the game. In addition, we can analyze whether the game is visually appealing and provides an acceptable immersive experience. 

Implementing an onboarding level is an essential step in developing a game to improve playability \cite{Cao2022LearningTutorials}. This level may be more crucial as players have less experience with HMDs and VR. Based on Lewis's benchmark for SUS \cite{Lewis2018ItemScale}, participants found the onboarding level easy to use (with an above-average benchmark). The simplicity of using the onboarding level is important as it is the first environment that users encounter when they enter VR. Considering that Project Beyond is an educational experience, learning the VR interactions and manipulating the game is our second priority. Therefore, it is important to provide an easy-to-follow and quick onboarding experience. The SUS results show that our tutorial level was successful. Our participants reported that they expect most people to learn to use the system very quickly, and they have confidence in using the system (with an above-average benchmark). 

In addition, we aimed to expose the participants to less demanding tasks in the tutorial level. Based on the NASA TLX results, we can observe that our onboarding level does not add extensive mental and physical demands. There was no time pressure during the tutorial level, so the temporal demand was very low. Additionally, without a high level of effort, they successfully accomplished the onboarding level without being frustrated. On the other hand, the escape room level is a challenging puzzle experience. Players must search, analyze, and think to solve the game's challenges. Accordingly, we observe high mental demand and effort at this level. Even considering the same interaction types in both levels, the physical demand is higher in the escape room. This could be due to the longer gameplay duration. 

Although the task load in the escape room level is relatively high, the user experience results show positive participant feedback. The high value for focused attention shows the potential of our environment to provide an engaging experience. Players will lose track of time and might spend more time in this environment, which can be beneficial for educational content. Additionally, participants enjoy the environment due to the attractive design and visual/auditory feedback.

\subsection{Limitations \& Future Work}
The study we present in this paper has several limitations: i) The number of participants was limited for the evaluation of this study. Extending the study with more participants can improve our understanding of the advantages and drawbacks of the environment. ii) We only considered VR and gaming experts as participants in our study. An evaluation with civil engineering students could lead to different results. iii) The study did not entail an evaluation of the learning effect itself. This is an important next step to validating our approach.

During the study, we realized that participants might enjoy a co-op experience to reduce mental demand and frustration. To test this theory, we will design an asymmetric co-op experience between VR and desktop clones of the game. This way, each user has a different point of view and functionality, but they can help each other to solve the puzzles.

This implementation, in general, has the potential to be extended to additional educational topics. Comparing this system's user experience and learning outcome with different learning contents would be interesting.

\subsection{Conclusion}
In this study, we designed and developed an immersive escape room game in Virtual Reality (VR). This gamified educational experience serves as a learning path for building energy simulation topics. Our approach of combining visual and audio hints with puzzles and educational content resulted in a highly immersive experience. However, it has the potential to be extended to other educational fields. The gameplay consists of a tutorial level to get used to VR interaction and the main puzzle room. To evaluate the usability of our implemented solution, we ran a user study with game development experts. The results are promising in the case of user engagement and usability. The evaluation shows that this VR game can increase user curiosity by providing a meaningful puzzle environment. Despite these positive responses, we found the task load relatively high for the escape room level. Extending the gameplay with a co-op scenario might reduce this issue and improve the user experience. 

\section*{Author Contributions}
\noindent
Game logic, data visualization, user study, API integration: G. A.-R.;
User interactions, user experience, tutorial: S. S.;
Room design: S. S. \& G. A.-R.;
Building energy simulation, educational content: M. G. \& C. H.;
Machine learning model, API: G. S.;
Supervision: J. P., C. H. \& G. S.;
Story: All authors.

\bibliographystyle{IEEEtran}
\bibliography{mendeley_import}

\end{document}